# Semiconductor-based selective emitter with sharp cutoff for thermophotovoltaic energy conversion

QING NI[*], RAJAGOPALAN RAMESH, CHENG-AN CHEN, LIPING WANG[*]

*School for Engineering of Matter, Transport & Energy, Arizona State University, Tempe, Arizona, 85287, USA*
*Corresponding author: qingni@asu.edu (Q.N.); liping.wang@asu.edu (L.W.)*

**Semiconductor emitter can possibly achieve sharp cutoff wavelength due to its intrinsic bandgap absorption and almost zero sub-bandgap emission without doping. A germanium wafer based selective emitter with front-side antireflection and backside metal coating is studied here for thermophotovoltaic (TPV) energy conversion. Optical simulation predicts the spectral emittance above 0.9 in the wavelengths from 1 to 1.85 μm and below 0.2 in the sub-bandgap range with sharp cutoff around the bandgap, indicating superior spectral selectivity behavior. This is confirmed by excellent agreement with indirectly measured spectral emittance of the fabricated Ge-based selective emitter sample. Furthermore, the TPV efficiency by paring the Ge-based selective emitter with a GaSb cell is theoretically analyzed at different temperatures. This work will facilitate the development of the semiconductor-based selective emitters for enhancing TPV performance.**

Thermophotovoltaic (TPV) devices convert thermal radiation from a high-temperature emitter to electricity via a narrow-bandgap photovoltaic (PV) cell. Since the emitter can be heated by any kind of heat source (combustible fuel, solar energy, waste heat, etc.), TPV technique has a wide range of applications. The theoretical efficiency of the TPV system has the Carnot limit [1]. However, due to the mismatch between the thermal radiation spectrum of the emitter and the absorption spectrum of the cell, the practical TPV system has low output electric power and poor efficiency [2-4]. To overcome this problem, much work has been carried out on spectrally selective emitters that emit photons with energy just above the bandgap of the PV cell. An ideal selective emitter should have an emissivity of one over a certain bandwidth just above the bandgap of the PV cell, and an emissivity of zero elsewhere. In addition, spectrally selective PV cells have also recently been demonstrated to enhance TPV efficiency exceeding 30% at 1455 K by recycling the unused photons to the emitter [5].

Previous works have demonstrated that the spectrally selective emitters can be achieved and tuned by photonic crystals [6-8] and metamaterials [9-12] structures. However, the complex geometries and the difficulty of fabrication process prevents the low-cost fabrication. High temperature stability after long-term practical operation with these nanostructures is another big concern. Spectrally selective emitters can also be realized by multilayer structures based on the anti-reflection effect or cavity resonance [4, 13, 14], which are much easier to fabricate potentially at large scale. Based on Kirchhoff's law of thermal radiation [15] that absorptivity equals emissivity at every wavelength in thermal equilibrium, spectrally selective absorbers may also serve as spectrally selective emitters. Wang et al. [13] previously reported a multilayer selective solar absorber which has good spectral selectivity behavior and high temperature stability. However, the cutoff wavelength, i.e., the transition between the high emittance above the bandgap and the low emittance below the bandgap, is not sharp, still resulting in some undesirable photons below the bandgap.

There is a less-studied category in semiconductor-based selective emitter, which has the potential for ideal emitter since the semiconductor bandgap edge itself provides a natural sharp cutoff wavelength. Previous works have theoretically designed semiconductor-metal tandem structures using Si and Ge mainly for solar thermal absorption [16, 17]. Tian et al. [18] experimentally demonstrated a low-cost single-crystal Si wafer based selective absorber with $Si_3N_4$ anti-reflective (AR) coating. Zhou et al. [19] fabricated a 9.6-μm thin-film Si based selective emitter with good mechanical flexibility and thermal stability at 600°C, and demonstrated strongly suppresses thermal emittance in the mid-IR over thicker silicon wafers. In addition, Oh et al. [20] proposed a selective emitter with a single layer of cylindrical structure of p-type silicon, which has a absorption peak in mid-IR range for TPV application. While most of the previous works on semiconductor-based selective absorber/emitter focused on Si, Ge which has a lower energy bandgap (0.67 eV) should be more suitable for TPV application.

In this work, we investigate a semiconductor-based selective TPV emitter made of a 500-μm-thick undoped Ge wafer with a 150-nm-thick $Si_3N_4$ front layer and a 200-nm-thick tungsten layer on the backside. The $Si_3N_4$ layer acts as an AR coating to reduce in-band reflection thus enhancing absorption or emission, while the tungsten layer is used to ensure the opaqueness of the structure with high reflectivity below the bandgap of Ge. In order to predict the spectral directional emittance, the indirect method is used here by calculating the spectral directional reflectance of the proposed structure. According to Kirchhoff's law, the spectral directional emittance is equal to the spectral directional absorptance as $1-R'_\lambda$ from energy balance with zero transmittance due to the opaque backside tungsten film. As illustrated in Fig. 1(a), $\rho_a$ or $\rho_b$ respectively represent the reflectivity of the Si$_3$N$_4$ thin layer for rays originated from air or from the Ge substrate, whereas $\tau_a$ or $\tau_b$ are the corresponding transmissivity. $\rho_s$ represents the reflectivity of the tungsten layer for rays originated from the Ge substrate. All above spectral directional reflectivity and transmissivity are calculated with the thin-film optics method [21, 22] for the Si$_3$N$_4$ front layer and the tungsten back coating separately. The light propagation inside the incoherent thick Ge substrate is taken into consideration by the ray-tracing method, and consequently, the spectral directional reflectance of the proposed Ge-based selective emitter with front and backside coatings can be calculated as [22]:

$$R'_\lambda = \rho_a + \rho_s \tau_a^2 \tau^2 / (1 - \rho_s \rho_b \tau^2) \quad (1)$$

where $\tau = \exp(-4\pi\kappa_s d_s / \lambda \cos\theta_s)$ is the internal transmissivity of the Ge substrate with extinction coefficient $\kappa_s$, thickness $d_s$, free-space wavelength $\lambda$, and refraction angle $\theta_s$ inside the Ge substrate calculated from incident angle $\theta_1$ with Snell's law.

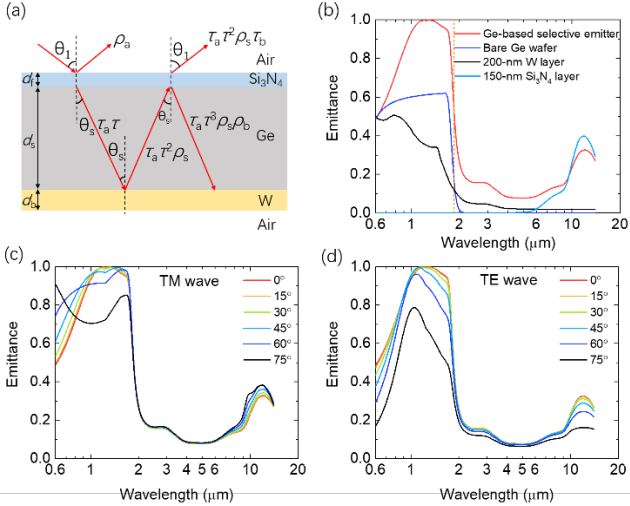

Fig. 1. (a) Schematic of the proposed Ge-based emitter structure and the wave propagation. (b) Theoretical normal spectral emittance of the proposed Ge-based emitter. Theoretical spectral emittance of the proposed Ge-based emitter at different incident angles for (c) TM wave and (d) TE wave.

Fig. 1 (b) shows the theoretical spectral emittance of the proposed Ge-based selective emitter. In the calculation, the optical constant of Si$_3$N$_4$ and tungsten are from Palik data [23], while the optical constants of undoped Ge are from Ref. [24]. For comparison, the simulated emittance spectra of a 150-nm Si$_3$N$_4$ layer, 200-nm tungsten film, and 500-μm bare undoped Ge wafer, were also shown in the figure where the dash line represents the Ge bandgap at 1.85 μm wavelength (i.e., 0.67 eV). As expected, undoped Ge has almost zero sub-bandgap emission with around 0.6 in-band emittance due to bandgap absorption. With the Si$_3$N$_4$ antireflection coating, the in-band emittance of the Ge-based selective emitter is significantly enhanced, reaching unity at wavelength of 1.3 μm right above the bandgap. Right below the bandgap, the emittance sharply drops to 0.1 or so due to the absorption of the lossy tungsten layer. At longer wavelengths beyond 10 μm, the emittance increases up to 0.3 or so because of strong phonon absorption of Si$_3$N$_4$. Fig. 1(c) and 1(d) respectively present the spectral emittance at different angles for transverse-magnetic (TM) and -electric (TE) polarized waves. It can be seen that at the wavelength below the bandgap, the emittance slightly increases with the incident angle for TM waves at first and then decreases at large incident angle, which is due to the Brewster effect [22, 25]. For non- or lightly absorbing materials, reflectance at TM waves could decrease to minimum at a particular incidence angle, namely Brewster angle. For the undoped Ge wafer used in this work, the Brewster angle is about 75° calculated by $\theta = \tan^{-1}(n_2/n_1)$ [22], where $n_2$ and $n_1$ represent the refractive index of Ge and air, respectively. Therefore, the reflectance at TM waves should decrease with incident angle from 0° to 75°, which explains the trend of emittance shown in Fig. 1(c). While for TE waves, the emittance decreases monotonically with the incident angle, which means the reflectance increases monotonically with the incident angle. Overall, the spectral emittance of the proposed structure is insensitive to the incident angles at small incident angles, indicating that the Ge-based selective emitter has diffuse emission behavior.

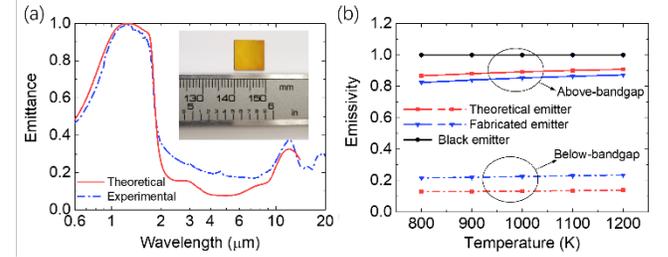

Fig. 2. (a) Spectral emittance of the fabricated Ge-based selective emitter sample. Inset is the photo of the fabricated Ge-based selective emitter on the Si$_3$N$_4$ side. (b) Spectrally averaged efficiency at different temperatures.

To experimentally demonstrate that the proposed structure has a good spectral selectivity, a Ge-based selective emitter sample was fabricated on a 500-μm-thick (100) double-sided polished Ge wafer (50 Ω·cm, MTI Corporation) with the size of 1×1 cm². Before the deposition of films, the Ge wafer was plasma cleaned. A 200-nm tungsten layer was firstly sputtered (Lesker PVD75 Sputter Coater) on one side of the Ge wafer at a rate of 1 Å/s under vacuum pressure of 1×10$^{-6}$ Torr. Then a 150-nm Si$_3$N$_4$ layer was deposited on the other side of the Ge wafer using Plasma-enhanced chemical vapor deposition (PECVD) method (Oxford Plasmalab100) at a rate of 290 Å/min under the temperature of 300°C. The root-mean-square surface roughness of the fabricated sample was found to be around 0.9 nm with atom force microscope.

The spectral normal reflectance of the fabricated sample was characterized by Fourier-transform spectroscopy (Thermo Fisher Scientific, Nicolet iS50) along with a specular reflection accessory (Harrick Scientific, Seagull) at an incidence angle of 8° in wavelength from 0.4 to 20 μm with a resolution of 4 cm$^{-1}$ in wavenumber. The spectral reflectance from 0.4 to 1 μm was measured by a Si detector, while that beyond 1 μm was measured by a deuterated triglycine sulfate (DTGS) detector. An aluminum mirror was used as the reference and the measured results was corrected by the theoretical reflectance of aluminum with optical constants obtained from Palik [23]. Each spectrum was averaged over 32 scans, and three independent measurements for both Al mirror and the sample were taken for consistency. The spectral normal emittance of the sample can be obtained by one minus spectral normal reflectance, as shown in Fig. 2(a). Inset is the photo of the fabricated sample. It can be observed that the measured spectral emittance of the fabricated sample agrees well with the simulated one above the bandgap. Below the bandgap the measured result is around 5% higher than the theoretical one, probably caused by the higher loss of the sputtered tungsten film due to grains and slightly oxidation that leads to different optical constants of tungsten from Palik.

To evaluate the spectral selectivity of the proposed emitter, we calculated the spectrally averaged emittance for above-bandgap spectrum $\bar{\varepsilon}_{\text{above}}$ and below-bandgap spectrum $\bar{\varepsilon}_{\text{below}}$ as:

$$\bar{\varepsilon}_{\text{above}} = \int_0^{\lambda_{\text{ge}}} \varepsilon_\lambda E_{b,\lambda}(T) d\lambda / \int_0^{\lambda_{\text{ge}}} E_{b,\lambda}(T) d\lambda \quad (2)$$

$$\bar{\varepsilon}_{\text{below}} = \int_{\lambda_{\text{ge}}}^\infty \varepsilon_\lambda E_{b,\lambda}(T) d\lambda / \int_{\lambda_{\text{ge}}}^\infty E_{b,\lambda}(T) d\lambda \quad (3)$$

where $\lambda_{\text{ge}} = 1.85$ μm is the bandgap wavelength of Ge, $E_{b,\lambda}(T) = 2\pi h c_0^2 / \lambda^5 [\exp(hc_0/\lambda k_B T) - 1]$ is the blackbody spectral emissive power [15] with Planck's constant $h$, the speed of light in vacuum $c_0$, Boltzmann constant $k_B$, and absolute temperature $T$. Fig. 2 (b) show $\bar{\varepsilon}_{\text{above}}$ and $\bar{\varepsilon}_{\text{below}}$ at different temperatures for both theortical Ge-based selective emitter and fabricated sample. It can be seen that both $\bar{\varepsilon}_{\text{above}}$ and $\bar{\varepsilon}_{\text{below}}$ of the theoretical Ge-based selective emitter barely change with the temperature between 800 and 1200 K. The same trend is shown for the fabricated sample. For the theoretical Ge-based selective emitter, the above-bandgap averaged emittance $\bar{\varepsilon}_{\text{above}}$ is 0.91 while the below-bandgap averaged emittance $\bar{\varepsilon}_{\text{below}}$ is 0.14 at 1200 K. In comparison, the black emitter has the emittance of 1 for the whole spectrum. Therefore, the proposed Ge-based selective emitter demonstrates good spectral selectivity.

The spectral efficiency can be defined as the percentage of the photons from the emitter absorbed by the PV cell:

$$\eta_{\text{spectral}} = \int_0^{\lambda_{\text{gc}}} q_{\text{e-c},\lambda} d\lambda / \int_0^\infty q_{\text{e-c},\lambda} d\lambda \quad (4)$$

where $\lambda_{\text{gc}}$ is the bandgap wavelength of the PV cell, $q_{\text{e-c},\lambda} = \dfrac{E_{be,\lambda}(T_e) - E_{bc,\lambda}(T_c)}{1/\varepsilon_{e,\lambda} + 1/\varepsilon_{c,\lambda} - 1}$ is the spectral net radiative heat flux between the emitter and the cell with a view factor of 1, and the subscript e(c) represents the emitter (cell). $\varepsilon_{e,\lambda}$ is the spectral emissivity of the emitter and $\varepsilon_{c,\lambda}$ is the spectral emittance (or absorptance) of the PV cell, both of which are assumed diffuse. $T_e$ and $T_c$ are respectively the emitter temperature and the cell temperature. GaSb cell with a close-to-Ge bandgap of 0.72 eV, which can be fabricated using molecular beam epitaxial and Zn diffusion methods to form p-n junction [26], is used here for the TPV modeling. $T_c$ is assumed to be 300 K in this work. Note that the in-band spectral absorptance of GaSb cell is obtained from Ref [26], and out-of-band absorptance is taken as 0.5 as a nominal value. Fig. 3(a) shows the spectral efficiency of the theoretical Ge-based selective emitter at different temperatures. For comparison, the spectral efficiencies of the fabricated sample and a black emitter ($\varepsilon_{e,\lambda}=1$) are also shown. The spectral efficiency will increase with the emitter temperature, which is because that the thermal radiation spectrum shifts to lower wavelength as the temperature increases, thus increasing the percentage of the photons with energies above the bandgap of the PV cell. Note that the theoretical Ge-based selective emitter has higher spectral efficiency than the fabricated sample due to its lower sub-bandgap emittance. The proposed Ge-based selective emitter has higher spectral efficiency than a black emitter. At a temperature of 1200 K, the spectral efficiency of the theoretical Ge-based selective emitter can achieve 34.0%, while that of a black emitter is only 12.0%.

To discuss the performance with the proposed Ge-based selective emitter, the TPV efficiency $\eta$ can be calculated by

$$\eta = P_e / q_{\text{in}} \quad (5)$$

where $P_e = J_{\text{sc}} V_{\text{oc}} FF$ is the maximum output electric power density produced by the PV cell. $J_{\text{sc}} = \int_0^{hc_0/E_g} \dfrac{e\lambda}{hc_0} \eta_{\text{IQE},\lambda} q_{\text{e-c},\lambda} d\lambda$ (A/cm$^2$) is the short-circuit current density [27]. Note that $E_g$ is the bandgap of the PV cell, $e$ is an elementary charge, and $\eta_{\text{IQE},\lambda}$ is the internal quantum efficiency (IQE) of the PV cell. $V_{\text{oc}} = (k_B T_c/e)\ln(J_{\text{sc}}/J_0 + 1)$ is the open-circuit voltage (V) [27], in which $J_0$ is the dark current calculated by $J_0 = e\left[n_i^2 D_h/(L_h N_D) + n_i^2 D_e/(L_e N_A)\right]$. $n_i$ is the intrinsic carrier concentration of the semiconductor, $N_D$ and $N_A$ are respectively the donor concentration and acceptor concentration, $D_h$ and $D_e$ are respectively the hole diffusion coefficient and electron diffusion coefficient, and $L_h$ and $L_e$ are respectively the hole and electron diffusion length. $FF$ is the filling factor calculated by $FF = (1 - 1/y)(1 - \ln y / y)$, in which $y = \ln(J_{\text{sc}}/J_0)$. All the calculation parameters of the GaSb cell are taken from Ref. [26]. $q_{\text{in}} = \int_0^\infty q_{\text{e-c},\lambda} d\lambda$ is the net radiative heat flux between the emitter and the cell with the same area.

Fig. 3(b), (c) and (d) respectively predict the TPV efficiency, the net radiative heat flux between the emitter and the PV cell, and the output power for the TPV system using the Ge-based selective emitter paired with a GaSb cell. For comparison, the results using a black emitter are also shown. With a black emitter at temperature from 800 to 1200 K, the TPV efficiency ranges from 0.3% to 4.0%. While with the fabricated Ge-based selective emitter sample, the TPV efficiency was improved ranging from 0.8% to 8.2%, which is due to the good spectral selectivity of the Ge-based selective emitter with low emittance below the bandgap, thus reducing the net radiative heat flux with almost the same power output. With a theoretical Ge-based selective

emitter, the TPV efficiency could be further improved ranging from 1.2% to 11.2%, which is due to its lower sub-bandgap emittance, thus further reducing the net radiative heat flux. Note that these three emitters produce similar $P_e$ below 1000 K. As the emitter temperature increases, the black emitter will surpass the proposed Ge-based selective emitter. $P_e$ depends on the net radiative heat flux above the bandgap of the cell. At low temperature, the emissive power from the emitter at short wavelength (above the bandgap) is low, which causes similar power from these three emitters. As the temperature increases, the thermal emission spectrum shifts to shorter wavelength. The black emitter has the highest emittance, resulting in the highest net radiative heat flux above the bandgap, thus producing the highest output power. While the fabricated Ge-based selective emitter sample has the lowest emittance, resulting in the lowest net radiative heat flux above the bandgap, thus producing the lowest power. In particular, at an emitter temperature of 1200 K, the theoretical Ge-based selective emitter can achieve a TPV efficiency of 11.2% and an output power of 2.3 kW/m².

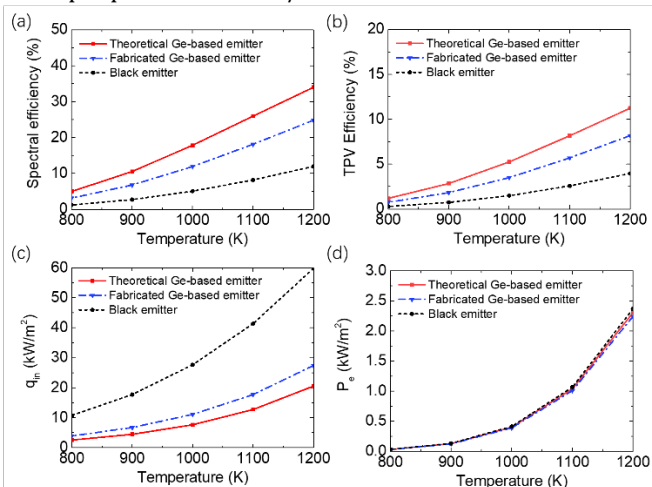

Fig. 3. (a) Spectral efficiency, (b) TPV efficiency, (c) net radiative heat flux and (d) output power at different temperatures from 800 K to 1200 K with theoretical Ge-based emitter, fabricated Ge-based emitter and black emitter paired with a GaSb cell.

In summary, a Ge wafer based selective emitter was proposed for thermophotovoltaic energy conversion. The optical property simulation of the proposed structure demonstrated a good spectral selectivity behavior. We also fabricated the Ge-based selective emitter sample and characterized its spectral emittance, which confirms the theoretical result. Moreover, the TPV system performance was simulated by using the proposed Ge-based selective emitter paired with a GaSb cell. The simulated results show a TPV efficiency of 11.2% and an output power of 2.3 kW/m² at 1200 K. This work will pave the way for the semiconductor-based spectrally selective emitter and will promote the development of the low-cost and high-efficiency TPV devices.

**Funding.** National Science Foundation (CBET-1454698); Air Force Office of Scientific Research (FA9550-17-1-0080).

**Acknowledgments.** We would like to thank Dr. Linshuang Long for taking the surface roughness measurement with AFM.

**Data Availability**. The data that support the findings of this study are available from the corresponding author upon reasonable request.

**Disclosures.** The authors declare no conflicts of interest.